\documentclass[11pt,showpacs,onecolumn]{revtex4}
\usepackage{amsmath}
\usepackage{amssymb,mathrsfs}
\usepackage{graphicx}
\usepackage{empheq}
\usepackage{graphicx,subfig}

\newcommand\ro{\hat\rho}
\newcommand\Ho{\hat H}
\newcommand\Ao{\hat A}
\newcommand\Bo{\hat B}

\newcommand\qo{\hat q}

\newcommand\so{\hat \sigma}
\newcommand\phio{\hat\phi}

\newcommand\Mc{\mathcal{M}}

\newcommand\Tr{\mathrm{Tr}}

\newcommand\half{\frac{1}{2}}

\begin{document}
\title{Exact Non-Markovian master equation for the Spin-Boson and Jaynes-Cummings models}   
\author{L. Ferialdi}
\email{ferialdi@fmf.uni-lj.si}
\affiliation{Department of physics, Faculty of Mathematics and Physics, University of Ljubljana, Jadranska 19, SI-1000 Ljubljana, Slovenia}
\date{\today}
\begin{abstract}
We provide the exact non-Markovian master equation for a two-level system interacting with a thermal bosonic bath, and we write the solution of such a master equation in terms of the Bloch vector. We show that previous approximated results are particular limits of our exact master equation. We generalize these results to more complex systems involving an arbitrary number of two-level systems coupled to different thermal baths, providing the exact master equations also for these systems. As an example of this general case we derive the master equation for the Jaynes-Cummings model.
\end{abstract}
\maketitle

Understanding the dynamics of a two-level system (TLS) coupled to an external environment is an ubiquitous problem in physics, chemistry and biology: quantum optics, charge transfer processes, tunneling phenomena, and light harvesting in photosynthetic systems are only few fields where the dissipative TLS covers a crucial role~\cite{BrePet02,Wang15}. The spin-boson model, i.e. a TLS interacting with a bosonic bath, is the paradigm for the description of these open systems~\cite{Legetal87,sb}. 
A first step in the understanding of the spin-boson model is given by the master equation in the Markovian approximation. The validity of this master equation is
restricted to those systems for which the environment can be assumed as static (i.e. the environment time scale is much shorter than that of the TLS). 
However, there are many processes where a Markov description is not sufficient~\cite{Leeetal}. In order to describe these systems one needs to consider a non-static bath, i.e. a bath that keeps track of the interaction with the TLS. Accordingly, some memory effects build up and the dynamics is non-Markovian. 
Several tentatives have been made to provide a non-Markovian master equation for the spin-boson model, exploiting e.g.  the noninteracting-blip approximation~\cite{Legetal87}, the time convolution-less technique (TCL)~\cite{BrePet02,BreKap01}, or the stochastic approach~\cite{stoch}. However, only approximated results were obtained.
The lack of an exact analytical description lead to investigate the problem by means of numerical techniques, among which we mention hierarchical equations of motion~\cite{heom}, quasi-adiabatic path integral~\cite{quapi}, effective modes~\cite{Cheetal14}, real-time RG in frequency space~\cite{rtrg1,rtrg2}, real-time FRG~\cite{rtrg2,rtfrg}, time-dependent DMRG~\cite{tddmrg}, and time-dependent NRG~\cite{tnrg,nrg}.

Another paradigmatic model is the (multimode) Jaynes-Cummings~\cite{jc}, which differs from the spin-boson only for the type of coupling between the TLS and the environment. This model is widely used in quantum optics and cavity-QED~\cite{Soletal}. Also the derivation of a non Markovian master equation for the Jaynes-Cummings model proved very difficult: an exact result has been obtained only for a bath in the ground state~\cite{BrePet02,Gar97}, while for a general thermal bath only approximated master equations are known.

In this Letter, we provide the solution of this long standing problem by deriving the exact (analytical) non-Markovian master equation for the spin-boson model, and by solving it in terms of the Bloch vector. Moreover, we provide the non-Markovian master equation for the Jaynes-Cummings model, and we extend our results to more complicated systems like the Tavis-Cummings model~\cite{TavCum68}, and the Jaynes-Cummings-Hubbard model~\cite{jch}.


The Hamiltonian of the spin-boson model can be written as follows~\cite{Legetal87}: $\Ho=\Ho_0+\Ho_I+\Ho_B$, where $\Ho_B$ is the Hamiltonian of the bath independent bosons, and $\Ho_0$ and $\Ho_I$ are respectively system and interaction Hamiltonians:
\begin{eqnarray}\label{Hsb}
\label{Hsb0}\Ho_0&=&-\half\Delta\hbar\so^x+\half\epsilon\so^z\,,\\
\label{HsbI}\Ho_I&=&\frac{k_0}{2}\so^z c^i \qo_i\,,
\end{eqnarray}
where $\so$ are Pauli matrices, $\qo_i$ are the positions of the bath oscillators, and $k_0,c^i$ are arbitrary real coupling constants. 
$\Delta$ and $\epsilon$ are respectively the detuning and dephasing constants (our result still holds if these are time dependent functions). For this reason, when $\Delta=0$ the model is called \lq\lq pure dephasing\rq\rq, and when $\epsilon=0$ is said \lq\lq pure detuning\rq\rq. The Einstein sum rule is understood.
We assume the initial state of the open system to be factorized, and the bosonic bath to be in a thermal state at temperature $T$. This can be fully characterized either by the environment spectral density $J(\omega)$, or by its hermitian two point correlation function $D(t,s)$, which are linked by well known expressions~\cite{BrePet02}:
\begin{eqnarray}
D^\mathrm{Re}(t,s)&=&\hbar\!\int_0^\infty \!\!d\omega J(\omega) \coth\!\left(\frac{\hbar\omega}{2k_B T}\right)\!\cos\omega(t-s),\\
D^\mathrm{Im}(t,s)&=&-\hbar\!\int_0^\infty \!\!d\omega J(\omega)\sin\omega(t-s)\,,
\end{eqnarray}
where $D^\mathrm{Re}$ and $D^\mathrm{Im}$ are  respectively real symmetric and imaginary antisymmetric parts of $D$, and $k_B$ is the Boltzmann constant.
 We introduce the left-right (LR) formalism \cite{Choetal85,DioFer14}, 
denoting by a subscript $L$ ($R$) the operators acting on $\ro$ from the left (right), 
e.g. $\Ao_{L}\Bo_{R}\ro=\Ao\ro\Bo$.
In a recent paper~\cite{DioFer14} it has been derived the most general trace preserving, completely positive, non-Markovian map $\Mc_t$, such that
$\ro_t=\Mc_t\ro_0$.
For a bilinear system-bath interaction of the type $\Ho_I= \Ao^i\phio_i$ (with $\Ao^i$ Hermitian system operators and $\phio_i$ Hermitian linear combinations of the bath modes), in interaction picture such a map reads:
\begin{eqnarray}\label{GMt}
\Mc_t\!&=&\!T\exp\bigg\{\int_0^t\!d\tau\!\!\int_0^t\!ds D_{jk}(\tau,s)\left[\!\Ao^k_{L}(s)\Ao^j_{R}(\tau)\!-\!\theta_{\tau s}\Ao^j_{L}(\tau)\Ao^k_{L}(s)\!-\!\theta_{s\tau}\Ao^k_{R}(s)\Ao^j_{R}(\tau)\!\right]\!\!\bigg\}\,,
\end{eqnarray}
where $\theta_{\tau s}$ denotes the step function that is $1$ for $\tau>s$, and the two-point correlation function is $D_{ij}=\Tr_B[\phio_i\phio_j\ro_B]$.  In the spin-boson interaction Hamiltonian~\eqref{HsbI}, the TLS is coupled to the environment via $\half k_0\so^z$. Hence, one just needs to define $\phio_z=c^i \qo_i$, and perform the substitution $\Ao\rightarrow\half k_0\so^z$ (there is only one $\Ao$) to obtain the correct map. After some manipulation, one finds that  the completely positive map describing the spin-boson model reads:
\begin{eqnarray}\label{sbMt}
\Mc_t\!&=&\!T\exp\left\{-\int_0^t\!d\tau\left[\so_{L}(\tau)-\so_{R}(\tau)\right]\int_0^\tau ds D(\tau,s)\so_{L}(s) - D^*(\tau,s)\so_{R}(s)\right\}\,,\nonumber
\end{eqnarray}
where the star denotes complex conjugation. In order to simplify the notation, we have dropped the index $z$, and we have absorbed the factor $\half k_0$ in $\so$. 
We observe that, by choosing a local correlation function $D(\tau,s)=D(\tau)\delta(\tau-s)$ one obtains the Markovian map~\cite{BrePet02}:
\begin{equation}\label{GMtL}
\Mc_t=T\exp\left\{\int_0^td\tau D(\tau)\left[\so_{L}(\tau)\so_{R}(\tau)-\hat{I}\right]\right\}\,,
\end{equation}
where $\hat{I}$ denotes the identity operator.
Differentiation of Eq.~\eqref{GMtL} provides the well known Lindbald equation. In order to obtain the non-Markovian master equation we need to differentiate the general $\Mc_t$ of Eq.~\eqref{sbMt}, and express $\dot{\Mc}_t$ in terms $\Mc_t$. This goal is hard to achieve because the double integral in the exponent of $\Mc_t$ is such that $\dot{\Mc}_t$ displays the time ordering of non-local arguments. This problem is overcame by exploiting the Wick's theorem~\cite{Wick}. We expand the map $\Mc_t$~\eqref{sbMt} in Dyson series:
\begin{equation}\label{Mseries}
\Mc_t=\sum_{n=0}^{\infty} \frac{(-1)^n}{n!}M^n_{t}\,,
\end{equation}
where $M^n_t=T\left[\prod_{i=1}^n\diamond_i\right]$, and
\begin{eqnarray}\label{Mn}
\diamond_i&=&\int_0^t\!dt_i\left[\so_{L}(t_i)-\so_{R}(t_i)\right]\int_0^{t_i} ds_i \left[D(t_i,s_i)\so_{L}(s_i)- D^*(t_i,s_i)\so_{R}(s_i)\right]\,.\nonumber
\end{eqnarray}
By differentiating $M^n_t$ one finds
\begin{eqnarray}\label{MDn}
\dot{M}_t^n=n \left[\so_{L}(t)-\so_{R}(t)\right]T\left[\!\int_0^{t} \!\!ds_1 \left(D(t,s_1)\so_{L}(s_1)\! -\! D^*(t,s_1)\so_{R}(s_1)\right)\!\prod_{i=2}^n \diamond_i\right].\nonumber
\end{eqnarray}
The main difference between $M^n_t$ and $\dot{M}_t^n$ is that the former are the time ordered products (T-products) of an even number of $\so$, while the latter display odd T-products. Here is where the Wick's theorem enters the calculations, allowing us to rewrite each $\dot{M}_t^n$ as a sum of even T-products, that are eventually rewritten in terms of $M^n_t$.
We note that different $\so$ acting on the same side of $\rho$ ($_{LL}$, $_{RR}$) anticommute with each other, while mixed contributions ($_{LR}$) commute:
\begin{equation}
\left\{\so_L,\so_L\right\}=\left\{\so_R,\so_R\right\}=0\,,\qquad \left[\so_L,\so_R\right]=0\,.
\end{equation}
Accordingly, a Wick contraction is defined as follows~\cite{Wick}:
\begin{eqnarray}
\label{contrLL} \hspace{-0.5cm}\overbracket{\so_L(s_1)\so_L(s_2)}\!=\! \overbracket{\so_R(s_1)\so_R(s_2)}\!&=&\! -\!\left\{\so(s_1),\!\so(s_2)\right\}\!\theta_{s_2,s_1},\\
\label{contrLR}  \overbracket{\so_L(s_1)\so_R(s_2)}&=&0\,.
\end{eqnarray}
Since $\Ho_0$ of Eq.~\eqref{Hsb0} gives linear Heisenberg equations for $\so^z$, these contractions are c-functions. This is a crucial feature because it implies that contractions commute with the T-ordering. 
Moreover, according to Eqs.~\eqref{contrLL},\eqref{contrLR} the contraction of two $\so$ separated by a product of $n$ $\so$ between them is
\begin{equation}
\overbracket{\so_L(s_1)(\dots)\so_L(s_2)}= (-1)^m \overbracket{\so_L(s_1)\so_L(s_2)}(\dots)\,,
\end{equation}
where $m\leq n$ is the number of $\so_L$ contained in $(\dots)$ (similarly for $R$ contractions).
These prescriptions allow us to rearrange the odd T-product of Eq.~\eqref{MDn} exploiting the Wick's theorem. Precisely, this is decomposed in an even T-product (that can be linked to $M_t^n$) plus another odd T-product of lower order with the same structure as the second line of Eq.~\eqref{MDn}. This procedure provides us with a rule that can we can apply recursively to $\dot{M}^n_t$, allowing us to decompose it in terms of even T-products. The calculations are rather involved and require some delicate manipulation. We report the details of the derivation in the Supplementary Material~\cite{sup}. The final result is the following integral master equation (the full notation has been restored):
\begin{eqnarray}\label{MEnl}
\dot{\ro}_t&=&-\frac{k_0^2}{4}\left(\so^z_{L}(t)-\so^z_R(t)\right)\left[\int_0^tds\mathbb{D}_{zz}(t,s)\so^z_L(s)-\mathbb{D}_{zz}^*(t,s)\so^z_R(s)\right]\ro_t\,,\nonumber
\end{eqnarray}
where 
\begin{equation}\label{DD}
\mathbb{D}_{zz}=\sum_{n=1}^\infty (-1)^{n-1} D_{zz(n)}\,.
\end{equation}
The explicit expressions of the $D_{zz(n)}$ are reported in~\cite{sup}.
The last step of our derivation is to provide a master equation that displays only operators at time $t$. We do so by solving the Heisenberg equations for $\Ho_0$: since these are linear we can write
\begin{equation}\label{soeasy}
\so^i(s)=b^i_{j}(s-t)\so^j(t)\,,
\end{equation}
where the indexes $i,j$ run over the components $x,y,z$ of the TLS, and $b$ is a real matrix (for explicit expressions of its entries see~\cite{sup}).
Substituting this expression in Eq.~\eqref{MEnl} one obtains
\begin{equation}\label{MEint}
\dot{\ro}_t=-\left(\so^z_{L}(t)-\so^z_R(t)\right)\left[B_{zi}(t)\so^i_L(t)-B^{*}_{zi}(t)\so^i_R(t)\right]\ro_t\,,
\end{equation}
with
\begin{equation}\label{Bzi}
B_{zi}(t)=\frac{k_0^2}{4}\int_0^tds\, \mathbb{D}_{zz}(t,s)b^z_{i}(s-t)\,.
\end{equation}
It is interesting to observe that the operators displayed by this master equation are: the coupling operator ($\so^z$), and the operators which are involved in the free evolution of the coupling operator ($\so^i$ through Eq.~\eqref{soeasy}). We further stress that Eq.~\eqref{MEint} has the same structure as the bosonic case~\cite{Fer16}: the difference among the two cases is encoded in the structure of the functions $B$. Moreover, in the weak coupling limit, these functions for the TLS and the bosonic case coincide as expected~\cite{BrePet02}.
Resorting to the Schr\"odinger picture and writing all the terms explicitly one eventually obtains
\begin{eqnarray}\label{MEsb}
\dot{\ro}_t&=&-i\left(\Ho_1(t)\ro_t-\ro_t\Ho_1^\dag(t)\right)-B^{Re}_{zz}(t)\left[\so^z,\left[\so^z,\ro_t\right]\right]\nonumber\\
&&+B_{zy}(t)\,\so^y\ro_t\so^z+B_{zy}^{*}(t)\,\so^z\ro_t\so^y+B_{zx}(t)\,\so^x\ro_t\so^z+B_{zx}^{*}(t)\,\so^z\ro_t\so^x\,,
\end{eqnarray}
where $\Ho_1(t)=\Ho_0+B_{zx}(t)\so^y-B_{zy}(t)\so^x$.
This is the exact non-Markovian master equation for the spin-boson model. We stress that all the functions displayed by this master equation are analytical. Moreover, if one chooses time dependent dephasing or detuning in $\Ho_0$, Eq.~\eqref{soeasy} still holds, and so does this master equation. The first line of Eq.\eqref{MEsb} displays a Lamb-shifted Hamiltonian and a dephasing term which changes only the non diagonal entries of $\ro_t$. The tunneling dynamics is driven by the second and third lines of Eq.~\eqref{MEsb}: these terms modify the populations of excited and ground states of the TLS. 
This master equation recovers, in the appropriate limits, the results known in the literature. For the full Hamiltonian~\eqref{Hsb0} the master equation for the spin-boson model is known in the weak coupling limit~\cite{Ines}: in this same limit (i.e. $\mathbb{D}=D$), our exact  master equation recovers that.
The only exact master equation known in the literature is the one for the \lq\lq pure dephasing\rq\rq~model, described by Eq.~\eqref{Hsb0} with $\Delta=0$. The master equation for this model is quite easy to derive because $\Ho_0$ and $\Ho_I$ commute. One can easily check that under this restriction $b^z_{x}=b^z_{y}=0$, $b^z_{z}=1$ and $\mathbb{D}=D$, that substituted in Eq.~\eqref{MEsb} lead to
\begin{equation}
\dot{\ro}_t=-i\frac{\epsilon}{2}\left[\so^z,\ro_t\right]-\frac{k_0^2}{4}\left(\int_0^t D(t,s)ds\right) \left[\so^z,\left[\so^z,\ro_t\right]\right]\,,
\end{equation}
which recovers the known master equation for this model~\cite{BrePet02,Guaetal14}.
Another interesting special case is the \lq\lq pure detuning\rq\rq model, i.e. Eq.~\eqref{Hsb0} with $\epsilon=0$. The master equation for this model is obtained simply by setting $B_{zx}=0$ in Eq.~\eqref{MEsb}. Such an exact master equation was not known, but if we restrict ourselves to the weak coupling limit 
we recover previously known approximated results~\cite{Sch07,CloBre12}. We further stress that Eq.~\eqref{MEsb} also provides the master equation for the Rabi model~\cite{BrePet02,rabi}: one simply needs to consider a \lq\lq one oscillator bath\rq\rq by taking a delta-correlated spectral density in Eqs.~(3),(4).

In order to solve Eq.~\eqref{MEsb} it is convenient to introduce the following identity:
\begin{equation}
\ro_t=\half\left(\hat{I}+\langle\sigma_i(t)\rangle\,\so^i\right)\,,
\end{equation}
where the vector  with components $\langle\sigma_i\rangle$ is known as Bloch vector. Substituting this equation in Eq.~\eqref{MEsb}, after some calculation one finds that the Bloch vector evolves according to the following equation:
\begin{equation}\label{sigmai}
\frac{d}{dt}\langle\sigma_i(t)\rangle=\mathcal{B}^{ij}(t)\langle\sigma_j(t)\rangle+\Sigma_i(t)\,,
\end{equation}
with $i,j=x,y,z$, $\Sigma=\left(-4 B^{Im}_{zy},\, 4 B^{Im}_{zx},\, 0\right)$, and
\begin{equation}\label{bloch}
\mathcal{B}=\left(
\begin{array}{ccc}
-4 B^{Re}_{zz}&-\epsilon&4 B^{Re}_{zx}\\
\epsilon&-4 B^{Re}_{zz}&4 B^{Re}_{zy}+\hbar\Delta\\
0&-\hbar\Delta&0
\end{array}\right)\,.
\end{equation}
This matrix recovers known results for the \lq\lq pure dephasing\lq\lq~and \lq\lq pure detuning\rq\rq~models~\cite{BreKap01}. However, unlike these special cases, the solution of the set of equations~\eqref{sigmai} with~\eqref{bloch} is non-trivial. In general, the dynamics of $\ro_t$ strongly depends on the bath spectral density and on the other parameters of the model. This important issue will be investigated in a dedicated forthcoming paper.
\begin{figure}
\begin{center}
\includegraphics[width=8.5cm]{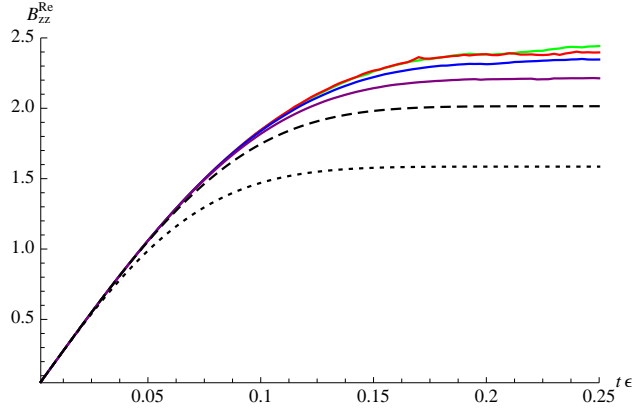}
\caption{Evolution of $B^{Re}_{zz}$ for increasing number of terms in the series~\eqref{DD} for $\mathbb{D}$. Dotted line is $n=1$, dashed $n=2$. Solid lines are respectively (bottom to top): $n=3$ (purple), $n=4$ (blue), $n=5$  (red), $n=6$  (green).
Bath with ohmic spectral density and Gaussian cutoff: $J(\omega)= 2\pi\omega \exp[-\omega^2\Lambda^{-2}]$. Other parameters are: $\epsilon=10$, $\Delta=\epsilon$,  $k^2_0=0.04 \epsilon$, $k_B T=0.1  \epsilon$, $\Lambda=2\epsilon$.} 
\end{center}
\end{figure}
Figure 1 shows the time evolution of $B^{Re}_{zz}(t)$ for an increasing number of terms in the series~\eqref{DD} for $\mathbb{D}$ (from $n=1$ to $n=6$). Black lines denote previously known results: dotted for weak-coupling limit (or second order TCL), and dashed for fourth order TCL (known for the \lq\lq pure detuning\rq\rq~ model only~\cite{BreKap01}). Colored (solid) lines are the original result of this Letter: the distance between the dashed line ($n=2$) and the green one (top solid line, $n=6$) clearly shows how previous results are improved. Moreover, besides small numerical errors, red ($n=5$) and green (two top solid) lines coincide, showing quite fast convergence of the series~\eqref{DD}. The evolution of the other coefficients of the master equation~\eqref{MEsb} display a similar convergence~\cite{sup}.

The method we presented can be exploited to obtain more general master equations. Indeed, the map~\eqref{GMt} provides the evolution for an interaction Hamiltonian of the type
\begin{equation}\label{HIgen}
\Ho_I= \so^i\phio_i
\end{equation}
(of which Eq.~\eqref{HsbI} is a special case). The superscript $i$ can be intended either as running over different TLSs, or as different components $(x,y,z)$ of the same system (or both these options). One can then repeat the calculations previously described, in the spirit of~\cite{Fer16}, and obtain the following master equation in interaction picture:
\begin{equation}\label{MEgen}
\dot{\ro}_t=-\left(\so^i_{L}(t)-\so^i_R(t)\right)\left[B_{ij}(t)\so^j_L(t)-B^*_{ij}(t)\so^j_R(t)\right]\ro_t\,,
\end{equation}
where we have to keep in mind that the correlation function has been promoted to a matrix $D_{ij}$, which implies
\begin{equation}\label{Bgen}
B_{ij}(t)=\int_0^tds\, \mathbb{D}_{ik}(t,s)b^k_{j}(s-t)\,.
\end{equation}
This exact non-Markovian master equation allows to describe many models of which only approximate master equations (or none) are known. Interesting examples falling in this category are 
the Tavis-Cummings model~\cite{TavCum68} and the Jaynes-Cummings-Hubbard model~\cite{jch}. We however stress that a crucial requirement is that the free Hamiltonian $\Ho_0$ must provide linear Heisenberg equations, otherwise Wick's contractions would not be c-functions (and the formalism would fail). Accordingly, spin-chains are excluded from our treatment~\cite{sup}.
We exploit this general result to attain the dynamics for the Jaynes-Cummings model which covers a fundamental role in the theories of quantum optics and cavity-QED~\cite{Soletal}. The interaction Hamiltonian for this model is obtained by applying the rotating wave approximation to Eq.~\eqref{HsbI}, and it reads
\begin{equation}\label{HIjc}
\Ho_I=\hbar g \left(\so^+ \sum_j \hat{a}_j+\so^- \sum_j \hat{a}^\dag_j\right)\,,
\end{equation}
where $\so^\pm=\so^x\pm i\so^y$. The free Hamiltonian for this model is $\Ho_0=\omega_0\so^+\so^-$ (our formalism allows to treat also the more general $\Ho_0$ of Eq.~\eqref{Hsb0}).
Since our formalism works with Hermitian operators, we rewrite this Hamiltonian as follows
\begin{equation}
\Ho_I=\frac{\hbar g}{2} \left[\so^x \sum_j \left(\hat{a}_j+\hat{a}_j^\dag\right)+\so^y \sum_j i\left(\hat{a}_j-\hat{a}_j^\dag\right)\right]\,.
\end{equation}
One observes that, although the Jaynes-Cummings coupling is an approximation of the standard spin-boson interaction~\eqref{HsbI}, it is of the general form~\eqref{HIgen}. We define $\phio_x=\sum_j \left(\hat{a}_j+\hat{a}_j^\dag\right)$ and $\phio_y=i\sum_j \left(\hat{a}_j-\hat{a}_j^\dag\right)$, and we exploit~\eqref{MEgen} to obtain
\begin{eqnarray}\label{MEjc}
\dot{\ro}_t&=&-i\left(\omega_0+B^{Re}_{xy}\right)[\so^+\so^-,\ro]+\left(B^{Re}_{xx}-B^{Im}_{xy}\right)\left(\so^-\ro\so^+-\half\{\so^+\so^-,\ro\}\right)\nonumber\\
&&+\left(B^{Re}_{xx}+B^{Im}_{xy}\right)\left(\so^+\ro\so^--\half\{\so^-\so^+,\ro\}\right)\,.
\end{eqnarray}
The new functions $B$ are defined by Eq.~\eqref{Bgen} and their expressions are analytical. One can check that if the bath is in its ground state (i.e. its temperature is zero), the following identity holds: $B^{Re}_{xx}=-B^{Im}_{xy}$, and Eq.~\eqref{MEjc} recovers the known master equation in this limit~\cite{BrePet02,Gar97,VacBre10}. In~\cite{VacBre10,SmiVac10} the authors provided an approximated master equation up to the fourth order TCL, for a larger class of initial bath states (namely those commuting with the number operator). Their master equation differs from ours by a \lq\lq dephasing\rq\rq~contribution of the type $\so^z\ro\so^z$. Equation~\eqref{MEjc} proves that such a contribution is null for thermal states. Precisely, a coupling of the type~\eqref{HIjc} will never display a contribution like $\so^z\ro\so^z$ because one of the two operators multiplying $\ro$ must always be the coupling operator (as explained after Eq.~\eqref{Bzi}). 
If one considers a more general $\Ho_0$ like that of Eq.~\eqref{Hsb0}, one obtains contributions of the type $\so^\pm\ro\so^z$, i.e. displaying at most one $\so^z$.

In this Letter, we provided the solution of a long standing problem, i.e. the exact non-Markovian master equation for the spin-boson model.  We solved such a master equation and we showed that our exact result recovers all known approximated results. Furthermore, we proved that the powerful formalism we developed allows to investigate more complicate systems that possibly involve more TLSs. As an example we provided the master equation for the Jaynes-Cummings model. 
Since the models investigated are the cornerstones for the analysis of more complicated systems, the results of this Letter will pave the way for new research on such systems, both under the analytical and numerical points of view.

\phantom{}
The author is indebted with A. Smirne for precious discussions and for providing fundamental help with the numerics of Fig.1. The author further thanks T. Prosen for useful conversations. 
This work was supported by the TALENTS$^3$ Fellowship Programme, CUP code J26D15000050009, FP code 1532453001, managed by AREA Science Park through the European Social Fund. 

\vspace{1cm}
\section*{\large Supplementary Material}

\noindent Although the idea underlying the derivation of the main result of this Letter is quite simple, the mathematics leading to it is delicate. 
The aim of this Supplementary Material is to guide the reader through the technical details of the calculations leading to the master equation~(20). Furthermore, we provide the plots showing the behavior of the functions displayed by the master equation~(20) and by the Bloch vector~(23).

\vspace{0.2cm}

\noindent {\bf Equations of motion and contractions.}

One can easily see that the Heisenberg equations of motion for the free Hamiltonian~(1) are:
\begin{eqnarray}\label{sys}
\dot{\so}^x&=&-\frac{\epsilon}{\hbar}\so^y\,,\nonumber\\
\dot{\so}^y&=&\frac{\epsilon}{\hbar}\so^x+\Delta\so^z\,,\\
\dot{\so}^z&=&-\Delta\so^y\,.\nonumber
\end{eqnarray}
Since this is a linear system, one can always find a unique solution, provided three boundary conditions. Since these can be freely chosen, we set them at time $t$, because they will be convenient to switch from interaction to Schr\"odinger picture. The solution of the system at any time $s\leq t$ reads 
\begin{equation}\label{soeasy2}
\so^i(s)=b^i_{j}(s-t)\so^j(t)\,,
\end{equation}
where
\begin{equation}
b(t)=\left(
\begin{array}{ccccc}
1+\frac{\epsilon^2}{\hbar^2\omega^2}(\cos\omega t-1)&&-\frac{\epsilon}{\hbar\omega}\sin\omega t&&\frac{\Delta\epsilon}{\hbar\omega^2}(\cos\omega t-1)\\\\
\frac{\epsilon}{\hbar\omega}\sin\omega t&&\cos\omega t&&\frac{\Delta}{\omega}\sin\omega t\\\\
\frac{\Delta\epsilon}{\hbar\omega^2}(\cos\omega t-1)&&-\frac{\Delta}{\omega}\sin\omega t&&1+\frac{\Delta^2}{\omega^2}(\cos\omega t-1)
\end{array}\right)\,,
\end{equation}
and $\omega^2=\Delta^2+\epsilon^2/\hbar^2$. We stress that if $\epsilon$ and $\Delta$ were time dependent functions, the system~\eqref{sys} would still be linear and it would still admit a solution of the type~\eqref{soeasy2}. Accordingly, our formalism can be applied also to time dependent detuning and dephasing. 

The standard definition of Wick contraction for spin 1/2 particles is
\begin{equation}
\overbracket{\so(s_1)\so(s_2)}= -\!\left\{\so(s_1),\!\so(s_2)\right\}\!\theta_{s_2,s_1}\,,
\end{equation}
where the unit step function $\theta$ is needed because we are not dealing with normal ordered products. One should also keep in mind that we have dropped the superscript $z$. In order to obtain the explicit expression of the contraction one simply needs to replace Eq.~\eqref{soeasy2} and exploit the anticommutation properties of the Pauli matrices. The result is
\begin{equation}
\overbracket{\so(s_1)\so(s_2)}= -2\big[b^z_{x}(s_1)b^z_{x}(s_2)+b^z_{y}(s_1)b^z_{y}(s_2)+b^z_{z}(s_1)b^z_{z}(s_2)\big]\theta_{s_2,s_1}\,.
\end{equation}
One now understands how crucial is to have linear equations of motion: only in this case one can write a solution in the form~\eqref{soeasy2} and obtain a contraction that is a c-function. If this is not the case, one cannot explicitly exploit the Wick's theorem and obtain the main result of this Letter. This explains why the formalism does not apply to spin chains. In fact, the equation of motion of e.g. the $x$ component of the $j$-th spin of a Heisenberg chain reads
\begin{equation}
\dot{\so}_j^x=-2 J_y \left(\so^z_j\so^y_{j+1}+\so^y_{j-1}\so^z_j\right)+2 J_z \left(\so^y_j\so^z_{j+1}+\so^z_{j-1}\so^y_j\right)\,,
\end{equation}
where $J_{y,z}$ are coupling constants displayed by the Hamiltonian.
An equation of this kind does not admit a solution of the type~\eqref{soeasy2}.

\vspace{0.2cm}
\noindent{\bf Calculation details leading to the master equation of Eq.(20).}
We start from the second line of Eq.(10) of the main text and we apply the Wick's theorem. For simplicity we focus only on the contribution from $\so_L$ (the calculations for $\so_R$ are similar). 
\begin{equation}\label{s7}
T\!\left[\left(\int_0^{t} \!\!ds_1 D(t,s_1)\so_{L}(s_1)\right)\prod_{i=2}^n \diamond_i\right]=\left(\int_0^{t} \!\!ds_1 D(t,s_1)\so_{L}(s_1)\right)T\!\left[\prod_{i=2}^n \diamond_i\right]+\int_0^{t} \!\!ds_1 D(t,s_1)\sum_i \overbracket{\so_L(s_1)T\left[ \prod_{i=2}^n \diamond_i\right]}\,,
\end{equation}
where $\diamond_i$ is given by Eq.(9).
Since all contractions contribute in the same way, we can rewrite the last term of Eq.~\eqref{s7} as follows:
\begin{eqnarray}\label{longcontr}
\lefteqn{(n-1)\int_0^tds_1 D(t,s_1)\Bigg(\overbracket{\so_L(s_1)\,T\Bigg[\int_0^{t} dt_2\int_0^{t_2} ds_2D(t_2,s_2)\big[\so_{L}(t_2)\so_{L}(s_2) - \so_{R}(t_2)\so_{L}(s_2)\big]}\prod_{i=3}^n \diamond_i\Bigg]}\\
&&\hspace{4cm}-\overbracket{\so_L(s_1)\,T\Bigg[\int_0^{t} dt_2\int_0^{t_2} ds_2\Big(D(s_2,t_2)\big[\so_{L}(t_2)\so_{R}(s_2) - \so_{R}(t_2)\so_{R}(s_2)\big]}\prod_{i=3}^n \diamond_i\Bigg]\Bigg)\,,\nonumber
\end{eqnarray}
where we simply extracted $\diamond_2$ from $\prod_{i=2}^n\diamond_i$, and the long overbracket denotes the contraction of $\so_L(s_1)$ with this term. We now exploit the rules of Eqs.(12)-(13) to express Eq.~\eqref{longcontr} in terms of single contractions as follows:
\begin{eqnarray}\label{}
\lefteqn{(n-1)\int_0^tds_1 D(t,s_1)\int_0^{t} dt_2\int_0^{t_2} ds_2\left[ \overbracket{\so(s_1)\so(t_2)}\big(D(t_2,s_2)\so_{L}(s_2)-D(s_2,t_2)\so_{R}(s_2)\big)\right.}\\
&&\hspace{6cm}\left.- \overbracket{\so(s_1)\so(s_2)}D(t_2,s_2)\big(\so_{L}(t_2)+\so_{R}(t_2)\big)\right]T\!\left[\prod_{i=3}^n \diamond_i\right]\,,\nonumber
\end{eqnarray}
where we also exploited the fact that contractions are c-functions and commute with T-ordering.
By manipulating the integral limit and rearranging the terms, one can rewrite this equation as follows:
\begin{eqnarray}\label{}
(n-1)\left(\int_0^tds_1 D(t,s_1)\int_0^{t} dt_2\int_0^{t} ds_2 \overbracket{\so(s_1)\so(t_2)}\big[\bar{D}(t_2,s_2)\so_{L}(s_2)-D^*(t_2,s_2)\so_{R}(s_2)\big]\right)T\!\left[\prod_{i=3}^n \diamond_i\right]\,,
\end{eqnarray}
where we have exploited the relation $D(s_2,t_2)=D^*(t_2,s_2)$ and we have defined
\begin{equation}
\bar{D}(t_2,s_2)=D^{Re}(t_2,s_2)(2\theta_{t_2s_2}-1)+iD^{Im}(t_2,s_2)\,.
\end{equation}
Repeating similar calculations for $\so_R(s_1)$ and recollecting the results, one eventually obtains:
\begin{eqnarray}\label{deco}
\lefteqn{T\!\left[\!\left(\int_0^{t} \!\!ds_1 D(t,s_1)\so_{L}(s_1)\! -\! D^*(t,s_1)\so_{R}(s_1)\right)\!\prod_{i=2}^n \diamond_i\right]=}\nonumber\\
&&\left(\!\int_0^{t} \!\!ds_1 D(t,s_1)\so_{L}(s_1)\! -\! D^*(t,s_1)\so_{R}(s_1)\right)T\!\left[\!\prod_{i=2}^n \diamond_i\right]\nonumber\\
&&+(n-1) T\!\left[\!\left(\!\int_0^{t} \!\!ds_1 D_{(2)}(t,s_1)\so_{L}(s_1)\! -\! D_{(2)}^*(t,s_1)\so_{R}(s_1)\!\right)\!\prod_{i=3}^n \diamond_i\right]\,,
\end{eqnarray}
with
\begin{equation}\label{D2}
D_{(2)}(t,s_1)=\int_0^tdt_2\int_0^tds_2\overbracket{\so(s_2)\so(t_2)}\big[\bar{D}(t_2,s_1)D(t,s_2)+D(t_2,s_1)D^*(t,s_2)\big]\,.
\end{equation}
The important lesson we learn from Eq.~\eqref{deco} is that the odd T-product of the left hand side, can be decomposed in an even T-product (that can be linked to $M_t^n$) plus another odd T-product of lower order with the same structure as the left hand side.
This implies that one just needs to perform the substitution $D\rightarrow D_{(2)}$ and repeat these calculations to obtain $D_{(3)}$, and so on. This iteration leads to the following expression:
\begin{equation}
D_{(n)}(t,s_1)=\int_0^tdt_n\int_0^tds_n\overbracket{\so(s_n)\so(t_n)}\left[\bar{D}(t_n,s_1)D_{(n-1)}(t,s_n)+D(t_n,s_1)D_{(n-1)}^*(t,s_n)\right]\,.
\end{equation}
The result of this procedure is that we have decomposed the initial odd T-product of Eq.~\eqref{s7} in a sum of even T-products, that can be linked to $M_t^k$:
\begin{eqnarray}\label{Tfin}
\lefteqn{T\!\left[\left(\!\int_0^{t} \!\!ds_1 D(t,s_1)\so_{L}(s_1)\! -\! D^*(t,s_1)\so_{R}(s_1)\!\right)\!\prod_{i=2}^n \diamond_i\right]=}\\
&&\hspace{2cm}\sum_{k=0}^{n-1}\frac{(n\!-\!1)!}{k!}\left(\!\int_0^{t} \!\!ds_1 D_{(n-k-1)}(t,s_1)\so_{L}(s_1)\! -\! D_{(n-k-1)}^*(t,s_1)\so_{R}(s_1)\!\right)M_t^k\nonumber.
\end{eqnarray}
By substituting this equation in Eq.~(10),
and by exploiting the definition of Cauchy product of two series one obtains 
\begin{equation}
\dot{\Mc}_t=-\sum_{n=1}^{\infty}(-1)^{n-1} \left[\so_{L}(t)-\so_{R}(t)\right] \left(\int_0^{t} ds_1 D_{(n)}(t,s_1)\so_{L}(s_1) - D_{(n)}^*(t,s_1)\so_{R}(s_1)\right)\sum_{k=0}^{\infty}\frac{(-1)^k}{k!}M_t^k\,.
\end{equation}
By applying this equation to $\ro_0$, one easily finds Eqs.(18),(19), where by definition $D_{(1)}\equiv D$, and we have added the subscript $zz$ for coherence with the notation of Eq.(5).

It is interesting to observe that Eq.~\eqref{D2} can be interpreted as the action of an operator $\mathfrak{D}$ on $D(t,s_2)$, i.e. $D_{(2)}(t,s_1)=\mathfrak{D}\left[D(t,s_2)\right]$, which for a general $D_{(n)}$ leads to $D_{(n)}(t,s_1)=\mathfrak{D}^{n-1}\left[D(t,s_2)\right]$.
According to this notation, one can rewrite Eq.(18) in a more elegant way, by formally summing the series:
\begin{equation}
\mathbb{D}(t,s_1)=\frac{1}{1+\mathfrak{D}}\left[D(t,s_2)\right]\,.
\end{equation}

The last step in the derivation of Eq.(20) requires the solution of the Heisenberg equations of motion for Eq.(1), which are provided in the first section of this Supplementary Material.


\vspace{0.2cm}
\noindent{\bf Plots of the functions in the master equation (20).}

In the main Letter, we provided a plot showing the time evolution of $B_{zz}^{Re}(t)$ for an increasing number of terms in the series (19) defining it. We provide here the plots for the functions $B_{zx}(t)$ and $B_{zy}(t)$, which rule the evolution of a density matrix according to Eq.(20). 
\begin{figure}[h!]
\begin{tabular}{ccc}
\subfloat[$B_{zx}^{Re}(t)$]{\includegraphics[width=7.5cm]{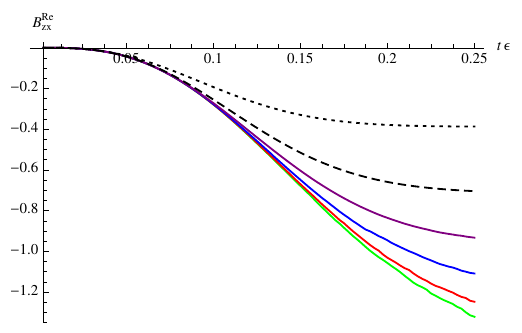}} 
  &\hspace{1cm} & \subfloat[$B_{zx}^{Im}(t)$]{\includegraphics[width=7.5cm]{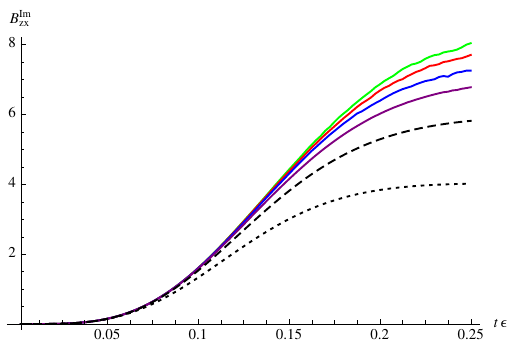}}\\
\subfloat[$B_{zy}^{Re}(t)$]{\includegraphics[width=7.5cm]{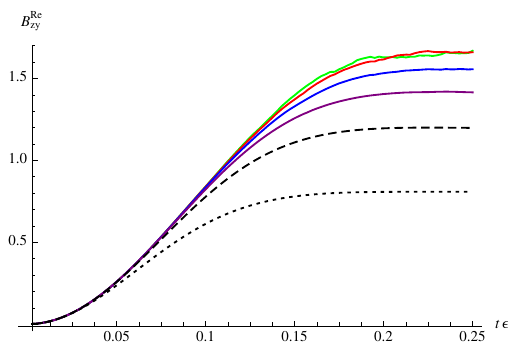}} 
  &\hspace{1cm}  & \subfloat[$B_{zy}^{Im}(t)$]{\includegraphics[width=7.5cm]{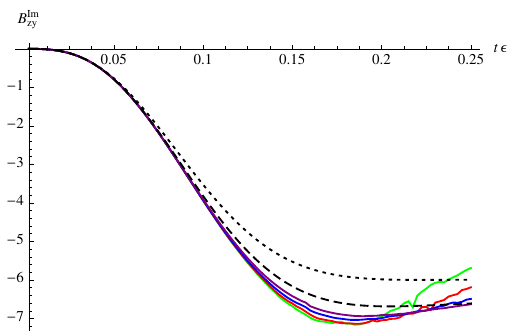}}\\
\end{tabular}
\caption{Dotted line is $n=1$, dashed $n=2$. Solid lines are respectively [bottom to top in insets (b) and (c), top to bottom in insets (a) and (d)]: $n=3$ (purple), $n=4$ (blue), $n=5$  (red), $n=6$  (green).
Bath with ohmic spectral density and Gaussian cutoff: $J(\omega)= 2\pi\omega \exp[-\omega^2\Lambda^{-2}]$. Other parameters are: $\epsilon=10$, $\Delta=\epsilon$,  $k^2_0=0.04 \epsilon$, $k_B T=0.1  \epsilon$, $\Lambda=2\epsilon$.}
\end{figure}

These plots clearly show how previous results (black lines) are improved, and that the series converges quite fast. We stress that these functions and the evolution of $\ro$ strongly depend on the bath structure and the other parameters of the model.

\newpage
\section*{ERRATUM}

In this paper, we have derived a master equation for two-level systems interacting with a bosonic bath. Such a master equation was claimed exact but, as we will show in this Erratum, this is not the case.
We start by correcting a typo in Eq. (11), that however does not play a role in the following considerations. In particular, different $\so$s acting on the same side of $\ro$ satisfy the standard anticommutation rules, and Eq. (11) should read:
\begin{equation}
\left\{\so_L,\so_L\right\}=\left\{\so_R,\so_R\right\}=2\hat{I}\,.
\end{equation}

The main result was derived starting from the most general completely positive, trace preserving, Gaussian, non-Markovian map~\cite{DioFer14}:
\begin{eqnarray}\label{GMt2}
\Mc_t\!&=&\!T_D\exp\bigg\{\int_0^t\!d\tau\!\!\int_0^t\!ds D_{jk}(\tau,s)\left[\!\Ao^k_{L}(s)\Ao^j_{R}(\tau)\!-\!\theta_{\tau s}\Ao^j_{L}(\tau)\Ao^k_{L}(s)\!-\!\theta_{s\tau}\Ao^k_{R}(s)\Ao^j_{R}(\tau)\!\right]\!\!\bigg\}\,,
\end{eqnarray}
that describes the reduced dynamics of a system (with operators $\Ao_j$) bilinearly interacting with a Gaussian bosonic bath (with correlation $D_{jk}(\tau,s)$). We have here added a subscript $_D$ to the time ordering operator $T$ in order to stress that this is the Dyson's (time) ordering, defined by:
\begin{equation}\label{TD}
T_D\left[\Ao_j(\tau)\Ao_k(s)\right]=\Ao_j(\tau)\Ao_k(s)\theta_{\tau s}+\Ao_j(s)\Ao_k(\tau)\theta_{s\tau}\,,
\end{equation}
where $\theta_{\tau s}=1$ for $\tau>s$, and zero elsewhere. We stress that Dyson's ordering is defined in the same way regardless of whether the system operators $\Ao$ are bosonic or fermionic.

The master equation was derived by applying Wick's theorem on the map~\eqref{GMt2}, according to which one can rewrite time-ordered products in terms of Wick contractions. However, Wick's theorem exploits an operator ordering that discriminates among bosons and fermions. Wick's (time) ordering $T_W$ coincides with Dyson's ordering  for bosons: $T_W=T_D$. Indeed, one can exploit Wick's theorem to obtain the exact master equation for a bosonic system~\cite{Fer16}. However, for fermionic systems the definition of Wick's ordering is different:
\begin{equation}\label{TW}
T_W\left[\Ao_j(\tau)\Ao_k(s)\right]=\Ao_j(\tau)\Ao_k(s)\theta_{\tau s}-\Ao_j(s)\Ao_k(\tau)\theta_{s\tau}\,
\end{equation}
In the main paper, this was accounted for both in the definition of contraction and in the explicit calculations. However, what was overlooked is that the map~\eqref{GMt2} is defined with $T_D$ also for fermionic systems: the derivation was performed by implicitly assuming that $\Mc_t$ was defined with $T_W$, i.e. $T_D$ was accidentally replaced by $T_W$ in Eq.~\eqref{GMt2}.

In order to solve this issue, one can conveniently express Dyson's ordering in terms of Wick's one (and vice versa). For the spin-boson model this amounts to:
\begin{equation}\label{TWTD}
T_W\left[\so^z(\tau)\so^z(s)\right]=T_D\left[\so^z(\tau)\so^z(s)\right]-2\so^z(s)\so^z(\tau)\theta_{s\tau}\,.
\end{equation}
This clearly shows that the replacement $T_D\rightarrow T_W$ implies to neglect systematically some contributions, leading to an approximate master equation. One might try to correct the previous derivation by taking these new terms into account. However, the relation~\eqref{TWTD} between the two orderings becomes very much complicated for higher order operator products $T\left[\so^z(\tau_1)\dots\so^z(\tau_n)\right]$, and one cannot rearrange such terms in order to obtain an overall contribution proportional to $\Mc_t$ (crucial to obtain a master equation; see Supplementary Material).
Another approach is to rewrite explicitly the product of Pauli matrices, by directly exploiting their properties. With our notation one finds e.g.
\begin{equation}\label{prod}
\so^z(\tau)\so^z(s)= {\bf b}^z(\tau)\cdot{\bf b}^z(s)\,\hat{I}+i({\bf b}^z(\tau)\times{\bf b}^z(s))\cdot{\boldsymbol \so}\,,
\end{equation}
where $\cdot$ and $\times$ denote respectively scalar and cross products, and bold symbols denote the vectors with components $x,y,z$. However, also in this case it is not possible to obtain a compact expression for higher order products.
The failure of these approaches is intimately connected with the algebra of Pauli matrices. In the first case, this is displayed by the fact that it is not possible to reduce a Dyson's ordering for fermions by means of the corresponding Wick's theorem. In the second approach, this is displayed by the terms proportional to ${\boldsymbol \so}$ in Eq.~\eqref{prod}, that do not allow one to reduce the complexity of higher order products. Although we have so far mentioned only the spin-boson model, a similar discussion applies to the master equation for the Jaynes-Cumming model.

A possible way out is the application of a so called \lq\lq generalized Wick's theorem\rq\rq~introduced in~\cite{WaShCa66}. This method however leads to multiply connected diagrams that might be difficult to recast in a master equation. An improvement in the result of this paper might be obtained by exploiting the path-integral formalism with time-non-local Lagrangians~\cite{FerBas12}.  Both these approaches require a detailed analysis and will be subject of further studies.

We have shown that our method systematically neglects some contributions in the derivation of the master equation, that accordingly cannot be considered exact. We however stress that the weak coupling limit of our master equation is correct and recovers known results. Indeed, this is obtained by neglecting the ordering of $\so(s_1)$ with the operators of $\prod_i\diamond_i$ in Eq. (10). Accordingly, $\so(s_1)$ is simply plugged out of the time ordering, bypassing the ordering issues described above.
On the other hand, Eq.~\eqref{prod} suggests that the exact master equation for the spin-boson model should include all combinations over $i,j=x,y,z$ of terms of the type $[\so_i,[\so_j,\ro]]$ and $[\so_i,\{\so_j,\ro\}]$. Similarly, the exact master equation for the Jaynes-Cummings model should display terms of the type $\so_z\ro\so_z$, that where missed by Eq. (30), but were obtained in~\cite{VacBre10,SmiVac10} by means of an approximated technique. 
One can in principle obtain an exact master equation, although this seems a very hard task, and as yet we could not find a closed expression.

\phantom{}
The author acknowledges M. Merkli for spotting the typo of Eq. (11), and M. Hall for useful comments. Furthermore, the author is indebted to G. Gasbarri for pointing out the ordering issue, and for collaboration in investigating the problem and drafting this Erratum.
This work was supported by the TALENTS$^3$ Fellowship Programme, CUP code J26D15000050009, FP code 1532453001, managed by AREA Science Park through the European Social Fund.

\end{document}